\documentclass[%
 reprint,
superscriptaddress,
 amsmath,amssymb,
 aps,
]{revtex4-2}

\usepackage{graphicx}
\usepackage{dcolumn}
\usepackage{bm}
\usepackage[T1]{fontenc}
\usepackage[utf8]{inputenc}
\usepackage{multirow}

\usepackage[linktocpage=true,
  colorlinks=true, 
  pdfborder={0 0 0},
  linkcolor=blue,
  citecolor=red,
  filecolor=yellow,
  urlcolor=blue,
  bookmarks,
  pdfauthor={},
]{hyperref}

\newcommand{\tc}{T$_{\text C}$}
\newcommand{\Tc}{T$_{\text C}$}

\begin{document}

\preprint{APS/123-QED}

\title{Ambient pressure high temperature superconductivity in RbPH$_3$ facilitated by ionic anharmonicity}
\author{{\DJ}or{\dj}e Dangi{\'c}}
	\email{dorde.dangic@ehu.es}
	\affiliation{Fisika Aplikatua Saila, Gipuzkoako Ingeniaritza Eskola, University of the Basque Country (UPV/EHU), 
		Europa Plaza 1, 20018 Donostia/San Sebasti{\'a}n, Spain}
	\affiliation{Centro de F{\'i}sica de Materiales (CSIC-UPV/EHU), 
		Manuel de Lardizabal Pasealekua 5, 20018 Donostia/San Sebasti{\'a}n, Spain}	
\author{Yue-Wen Fang}
	\email{yuewen.fang@ehu.eus}
	\affiliation{Fisika Aplikatua Saila, Gipuzkoako Ingeniaritza Eskola, University of the Basque Country (UPV/EHU), 
		Europa Plaza 1, 20018 Donostia/San Sebasti{\'a}n, Spain}
	\affiliation{Centro de F{\'i}sica de Materiales (CSIC-UPV/EHU), 
		Manuel de Lardizabal Pasealekua 5, 20018 Donostia/San Sebasti{\'a}n, Spain}	
\author{Tiago F. T. Cerqueira}
    \affiliation{CFisUC, Department of Physics, University of Coimbra, Rua Larga, 3004-516, Coimbra, Portugal}
\author{Antonio Sanna}
\affiliation{Max-Planck-Institut für Mikrostrukturphysik, Weinberg 2, D-06120, Halle, Germany}
\author{Miguel A. L. Marques}
\affiliation{Research Center Future Energy Materials and Systems of the University Alliance Ruhr and Interdisciplinary Centre for Advanced Materials Simulation, Ruhr University Bochum, Universitätsstraße 150, D-44801, Bochum, Germany}
\author{Ion Errea}%
	\email{ion.errea@ehu.eus}
	\affiliation{Fisika Aplikatua Saila, Gipuzkoako Ingeniaritza Eskola, University of the Basque Country (UPV/EHU), 
		Europa Plaza 1, 20018 Donostia/San Sebasti{\'a}n, Spain}
	\affiliation{Centro de F{\'i}sica de Materiales (CSIC-UPV/EHU),
		Manuel de Lardizabal Pasealekua 5, 20018 Donostia/San Sebasti{\'a}n, Spain}
	\affiliation{Donostia International Physics Center (DIPC),
		Manuel de Lardizabal Pasealekua 4, 20018 Donostia/San Sebasti{\'a}n, Spain}

\date{\today}

\begin{abstract}
Recent predictions of metastable high-temperature hydride superconductors give hope that superconductivity at ambient conditions is within reach. In this work, we predict RbPH$_3$ as a new compound with a superconducting critical temperature around 100 K at ambient pressure, dynamically stabilized thanks to ionic quantum anharmonic effects. RbPH$_3$ is thermodynamically stable at 30 GPa in a perovskite $Pm\bar{3}m$ phase, allowing its experimental synthesis at moderate pressures far from the megabar regime. 
With lowering pressure it is expected to transform to a $R3m$ phase that should stay dynamically stable thanks to quantum fluctuations down to ambient pressures. Both phases are metallic, with the $R3m$ phase having three distinct Fermi surfaces, composed mostly of states with phosphorus and hydrogen character. The structures are held together by strong P-H covalent bonds, resembling the pattern observed in the high-temperature superconducting H$_3$S, with extra electrons donated by rubidium. 
These results demonstrate that quantum ionic fluctuations, neglected thus far in high-throughput calculations, can stabilize at ambient pressure hydride superconductors with a high critical temperature.
\end{abstract}

\maketitle



Recently discovered high-pressure hydrides have demonstrated record-breaking superconducting critical temperatures (T$_\mathrm{C}$'s)~\cite{hydrides1, hydrides2, hydrides3, hydrides4, FLORESLIVAS20201}. Despite the fact that the first discoveries happened only less than a decade ago, this research area has already produced a multitude of distinct compounds exhibiting superconductivity at temperatures exceeding that of liquid nitrogen~\cite{H3S, LaH10, YH6, YH96, CaH6, LaBeH8}. First-principles calculations played a leading role in this unprecedented success, predicting beforehand some of the materials that were later experimentally confirmed as high-temperature superconductors~\cite{H3Spred, PengPRL, LiuPNAS, wang_superconductive_2012, LaBH8}. While the technological applications of these materials are potentially boundless, the necessity of large pressures for stabilizing them hinders their practical implementation.

For this reason, the community's focus shifted towards searching for polyhydride materials that can be synthesized at relatively low pressures, below the megabar regime~\cite{MgIrH6_ion,MgIrH6_chris,hydrides_at_ambient1}. The hope is that systems that are thermodynamically stable, and thus synthesizable, at lower pressures are more likely to be metastable even at ambient conditions. This approach has already yielded several promising candidates. For example, two separate works proposed Mg$_2$IrH$_6$, with a T$_\mathrm{C}$ between 80--160 K~\cite{MgIrH6_chris, MgIrH6_ion}. Dolui et al. find that the structure is metastable at 0 GPa, with thermodynamical stability reached at 15 GPa, which is the proposed synthesis pressure~\cite{MgIrH6_chris}. Similar conclusions were reached in Ref.~\cite{MgIrH6_ion}, marking this finding as an important milestone in the search for superconductivity at ambient conditions.

Very recently there were several works suggesting perovskite hydrides as a feasible route towards ambient condition superconductivity~\cite{hydrides_at_ambient1, perovskite1, perovskite2, perovskite3}. Using a machine learning and high-throughput search, it has been suggested that several perovskite hydrides could host superconductivity, both normal and inverted single and double perovskites, as well as perovskite derivatives of SM$_2$-TM-H$_6$ stoichiometry~\cite{hydrides_at_ambient1}. None of the suggested structures are found to be on the convex hull, i.e. all of them are thermodynamically unstable at 0 GPa. On the other hand, several promising candidates have been described in the single perovskite family, thermodynamically and dynamically stable below 50 GPa~\cite{perovskite1}. Finally, another high-throughput study~\cite{perovskite3} suggested the perovskite KAlH$_3$ compound as a possible superconductor at ambient pressure. However, the authors do not comment on the thermodynamical stability of this structure and it is an open question whether this material is likely to be synthesized. 

All previous high-throughput searches were performed discarding structures that are unstable in the harmonic approximation, i.e. those that are not a minimum of the Born-Oppenheimer energy surface (BOES) but a saddle point. However, it is now clear that quantum and anharmonic effects of the ions promote the stability of superconducting phases of high-pressure hydrides~\cite{H3S_ion, H3S_ion2, LaH10_ion, LuNH1, Lu4H11N_ion}. Hydrogen atoms have a very low ionic mass and thus a large mean square displacement even at zero temperature, allowing them to explore the anharmonic parts of the BOES. This can significantly change the lowest free energy structure as well as the vibrational characteristics of the material, having a crucial role in its superconducting properties. 
The inclusion of quantum and anharmonic effects usually promotes higher symmetry structures at lower pressures~\cite{H3S_ion, LaH10_ion} by improving both dynamical and thermodynamical stability. This fact is always overlooked in the high-throughput screening studies for new hydride superconductors and represents a possible new pathway toward ambient condition superconductivity. 

By making use of \emph{ab initio} calculations including ionic quantum fluctuations and anharmonicity within the stochastic self-consistent harmonic approximation (SSCHA)~\cite{SSCHA1, SSCHA2,SSCHA3,SSCHA4,SSCHA5}, here we predict a novel RbPH$_3$ compound with a critical temperature at ambient pressure around 100 K. By studying the ternary Rb-P-H phase diagram, we reveal that at 25 GPa the $R3m$ perovskite-like structure of RbPH$_3$ is very close to the convex hull, with a decomposition path to P, RbPH$_2$, and Rb$_2$PH$_9$. Including the vibrational contribution to the energy leads to the stabilization of this structure with respect to the decomposition path already at 30 GPa, making it synthesizable at moderate pressures.
Quantum and anharmonic effects make this structure metastable down to ambient pressure, indicating that these effects can lead to the stabilization of many high-T$_{\mathrm{C}}$ hydrides down to ambient pressure.

\begin{figure}[t!]
    \centering
    \includegraphics[width=1.0\linewidth]{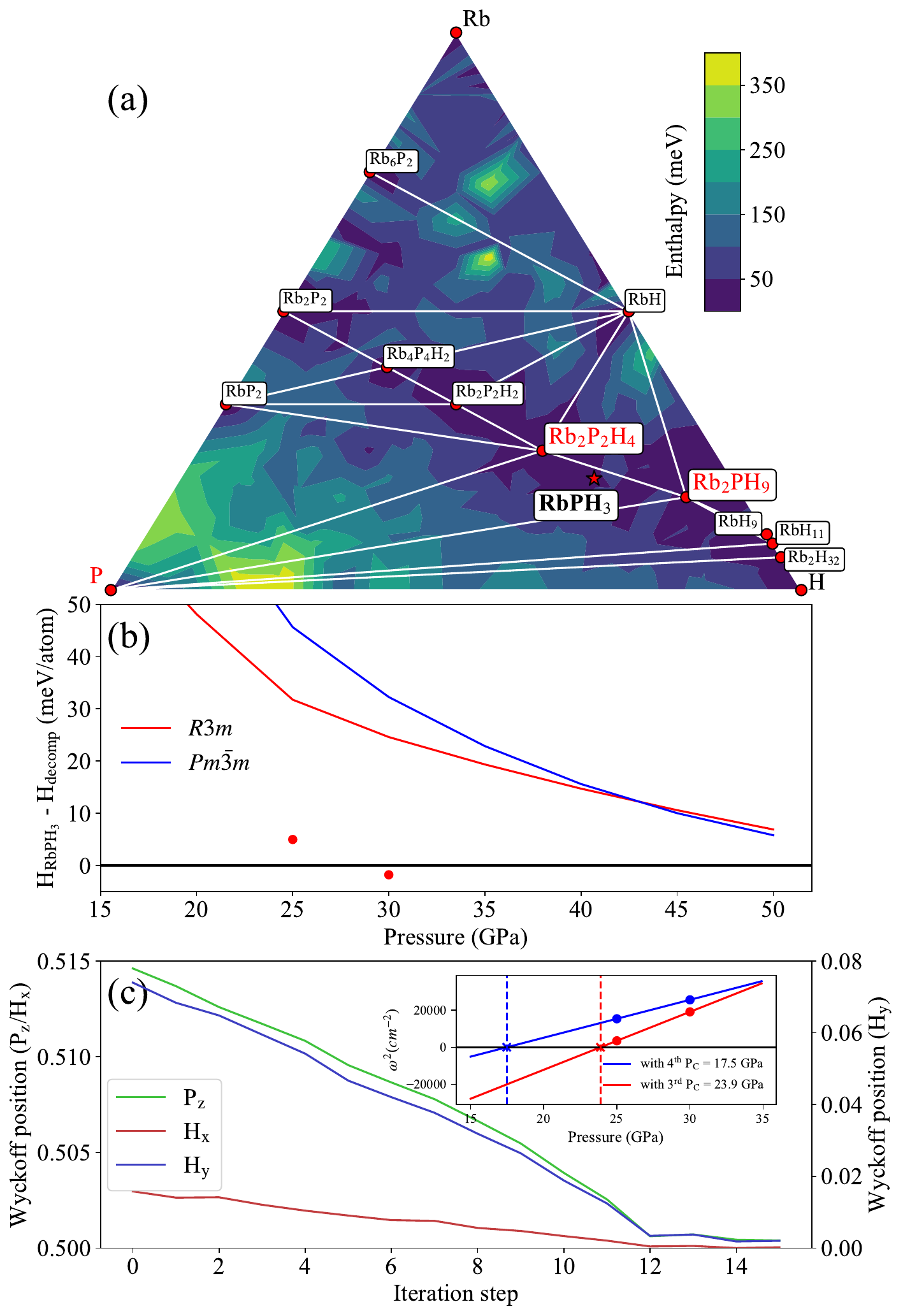}
    \caption{(a) Ternary phase diagram for the Rb-P-H system neglecting the zero-point energy at 25 GPa. Stoichiometries on the convex hull are represented with red circles. RbPH$_3$ is denoted by the star. The color scheme of the background denotes the enthalpy from the convex hull. (b) Continuous lines are enthalpy differences between RbPH$_3$ phases against the decomposition path neglecting the zero-point energy. The points are the same differences with the inclusion of the zero-point energy at the harmonic level. (c) The change of free Wyckoff positions of the $R3m$ phase during the SSCHA minimization at 25 GPa. The inset shows the pressure dependence of $\Gamma _4 ^{-}$ phonon of $Pm\bar{3}m$ phase that drives the $Pm\bar{3}m \rightarrow R3m$ transition calculated with the SSCHA including 3$^{\mathrm{rd}}$ and $4^{\mathrm{th}}$ order anharmonicity (see Supp. Mat.~\cite{supp_mat}).}
    \label{fig:phasediagram}
\end{figure}

Rb-H has been reported as a promising binary system for high-temperature superconductivity at relatively low pressures, around 50 GPa~\cite{RbH}. The hope is that by adding a third element, high-\tc{}  structures at even lower pressures can be stabilized. To check this we constructed a phase diagram for the Rb-P-H system at 25 GPa calculating the enthalpies without considering the energy associated with the ionic quantum fluctuations, in other words, assuming that the total energy is given just by the BOES (see Fig~\ref{fig:phasediagram} (a)). We explored 258 stoichiometries sampled from Alexandria~\cite{Alexandria1, Alexandria2}, Super-Hydra~\cite{SuperHydra}, and Deepmind~\cite{deepmind} databases, and from \emph{ab initio} global structure predictions using the Minima Hopping Method~\cite{Goedecker2004,Amsler2010} (for details, please refer to the Supp. Mat.~\cite{supp_mat}). 
We find that at 25 GPa there are several different compounds on the convex hull: RbP$_2$, RbH$_9$, RbPH, Rb$_3$P, Rb$_2$P$_2$H, RbPH$_2$, RbP, Rb$_2$PH$_9$, RbH, RbH$_{11}$, and RbH$_{16}$. Most of these structures have a low symmetry and are insulators, and thus not relevant in the quest for high-T$_\mathrm{C}$ superconductors. Nevertheless, one of the stoichiometries that is close to the convex hull is the metallic perovskite RbPH$_3$, which is at a distance of merely 33 meV/atom from the convex hull.  

The most probable decomposition path of RbPH$_3$ is found to be RbPH$_2$ + P + Rb$_2$PH$_9$. Since the energy differences are very small (33 meV/atom), we decided to check whether the zero-point energy is enough to stabilize thermodynamically the $R3m$ phase. We calculated harmonic phonons for all four phases ($R3m$ RbPH$_3$, RbPH$_2$, P, and Rb$_2$PH$_9$) and calculated the zero-point energy for each structure. Including the harmonic vibrational energy at this pressure reduces the excess enthalpy to only 6 meV/atom. Already at 30 GPa, the enthalpy of the $R3m$ phase is below the convex hull (3 meV/atom). This means that at this pressure we can expect the synthesis of the material at low temperatures. In the harmonic approximation, increasing the temperature makes RbPH$_3$ even more thermodynamically stable, with 7 meV/atom at 1000 K below the decomposition path. Therefore, the pressure needed to synthesize this material will be lower at elevated temperatures, facilitating its synthesis with laser heating methods. 

\begin{figure}[t!]
    \centering
    \includegraphics[width=1.0\linewidth]{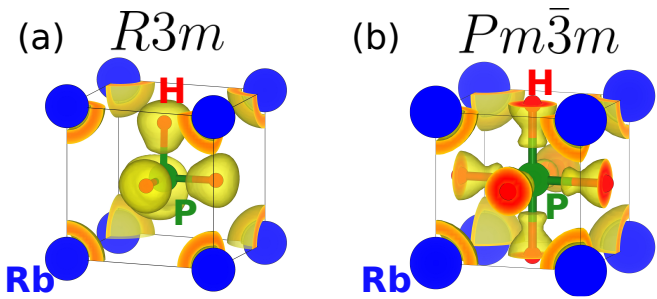}
    \caption{Structure of (a) the $R3m$ phase at 0 GPa and (b) the $Pm\bar{3}m$ phase at 30 GPa of RbPH$_3$. The contours show the electronic localization function (ELF) at the value of 0.8. The visualization of the structure and the ELF is done using VESTA~\cite{VESTA}.}
    \label{fig:elf}
\end{figure}

Once the thermodynamic stability of the $R3m$ phase of RbPH$_3$ has been determined, we proceed by considering the impact of ionic quantum fluctuations and anharmonicity on this structure by relaxing it at 25 GPa and 100 K with the SSCHA (see Fig.~\ref{fig:phasediagram} (c)). In the $R3m$ phase, Rb and P atoms occupy the 3a Wyckoff sites, whose representative can be written in the rhombohedral description as $(z,z,z)$ and the H atoms occupy the 9b sites, whose representative is $(x,y,x)$. At 25 GPa the rhombohedral angle at the classical BOES level is very close to a cubic one (90.01$^{\circ}$) and does not change significantly during the SSCHA minimization. The SSCHA minimization, however, strongly impacts the free parameters in the structure, by shifting them away from the BOES minimum to reach $z_{\mathrm{Rb}}=0$, $z_{\mathrm{P}}=0.5$, $x_{\mathrm{H}}=0.5$, and $y_{\mathrm{H}}=0.0$. These Wyckoff parameters correspond to those of the higher symmetry $Pm\bar{3}m$ space group, as after the SSCHA relaxation Rb atoms are at the 1a site, P atoms at the 1b site, and H atoms at the 3c sites, which are the positions of the simple cubic perovskite structure (see Fig.~\ref{fig:elf} for the crystal structures). All this suggests that, according to SSCHA, the actual free energy minimum that considers ionic quantum fluctuations at 25 GPa and 100 K is the $Pm\bar{3}m$ perovskite phase and not the $R3m$ phase, which despite being at the minima of the BOES is no longer a local minimum of the total free energy.

\begin{figure}[t!]
    \centering
    \includegraphics[width=0.5\textwidth]{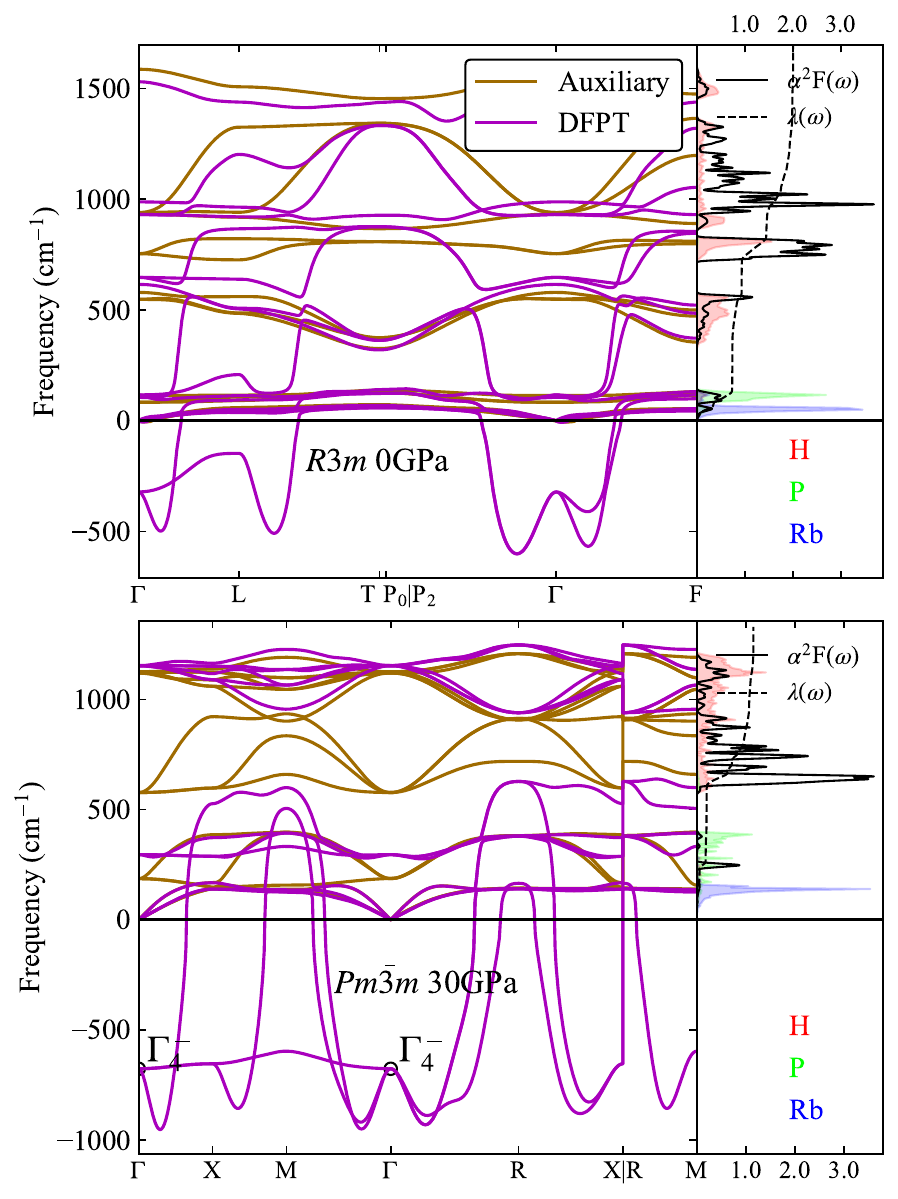}
    \caption{Phonon band structures of RbPH$_3$ in the $R3m$ phase at 0 GPa and the $Pm\bar{3}m$ phase at 25 GPa calculated at the harmonic level with DFPT and the auxiliary SSCHA force constants. Both calculations are done on the structure that is the minima of the total free energy (the result of SSCHA minimization). The side plot shows the Eliashberg spectral function $\alpha ^2 F(\omega)$ and the electron-phonon coupling constant $\lambda (\omega)$. The colour-coded background is the atom-resolved phonon density of states calculated from the SSCHA result. We scaled the phonon density of states in order to represent it in the same graph as $\alpha ^2F(\omega)$.}
    \label{fig:phonons}
\end{figure}

The harmonic phonon calculations of the $Pm\bar{3}m$ phase at 25 GPa and 30 GPa reveal large phonon instabilities throughout the whole Brillouin zone, in particular, a triply degenerate $\Gamma _4 ^-$ phonon mode (see Fig.~\ref{fig:phonons}). 
Group theory recognizes this mode as the one responsible for the $Pm\bar{3}m \rightarrow R3m$ phase transition. The SSCHA relaxation of the structure is able to stabilize this phonon mode at these two pressures, demonstrating that the $Pm\bar{3}m$ phase is stabilized by quantum anharmonic effects. By extrapolating the square of the SSCHA phonon frequency of the $\Gamma _4 ^-$ phonon mode to lower pressures (see inset of Fig.~\ref{fig:phasediagram} (c)), we estimate that the $Pm\bar{3}m$ phase undergoes a second-order phase transition to the $R3m$ close to 20 GPa. Below this pressure, the $R3m$ remains the ground state on the free energy landscape down to 0 GPa. The SSCHA relaxation at ambient pressure for this phase yields a structure with a rhombohedral angle of $91.1^{\circ}$ and Wyckoff positions that clearly deviate from the cubic values (see Supp. Mat.~\cite{supp_mat}). The harmonic phonons of this phase reveal clear instabilities in the vicinity of $\Gamma$, which the SSCHA is able to correct with the inclusion of 4$^{\mathrm{th}}$ order terms in the calculation of the Hessian of the free energy. This confirms the dynamical stability of this phase at ambient pressure (for a more detailed analysis and computational details see Supp. Mat.~\cite{supp_mat}). Since the SSCHA considers the kinetic energy of the ions in the calculation of the free energy, it also suggests that the $R3m$ phase of RbPH$_3$ is kinetically stable at ambient pressure and, thus, metastable. It is worth remarking that, since at the quasiharmonic level the vibrational contribution to the energy reduces the pressure to make RbPH$_3$ thermodynamically stable, the inclusion of anharmonic effects in the calculations of the convex hull would make it even more favorable and reduce even more the pressure needed to synthesize it.




In order to understand the bonding nature in RbPH$_3$, in Fig.~\ref{fig:elf} we show the electronic localization function (ELF) of the $R3m$ and $Pm\bar{3}m$ phases at 0 and 30 GPa, respectively. The ELF reveals a strongly covalently bonded PH$_3$ cluster in the $R3m$ phase, with strong P-H covalent bonds, which are connected among each other through weaker hydrogen bonds. Rb atoms stay, on the contrary, isolated. This is not surprising given the trivalency of P. At higher pressures in the $Pm\bar{3}m$ phase the covalent and hydrogen bonds symmetrize forming the octahedron of the perovskite structure. The bonding pattern clearly resembles the one in the high-T$_{\mathrm{C}}$ H$_3$S superconductor with $Im\bar{3}m$ symmetry~\cite{H3Spred,networkingvalue}. In both cases, the Bader charge analysis shows a small charge transfer between P/S and H atoms in the cubic phase. 
The only difference is that in $Pm\bar{3}m$ RbPH$_3$, it is Rb (Bader charge-based ionic charge of +0.727$e$) that donates electrons both to phosphorus (-0.118$e$) and hydrogen (-0.203$e$). On the other hand, in H$_3$S, hydrogen has an ionic charge of +0.066$e$ and sulfur -0.199$e$.
The extra charge donated by Rb promotes the $Pm\bar{3}m$ phase with symmetric P-H-P bonds at lower pressure. This clearly shows that trivalent atoms doped with alkali metals are likely to form metallic perovskite hydrides with strong covalent bonds.


Figure~\ref{fig:phonons} shows the vibrational properties of the $R3m$ and $Pm\bar{3}m$ phases calculated at the harmonic level with density functional perturbation theory (DFPT)~\cite{baroni_phonons_2001} and at the anharmonic level with the SSCHA for the structures that minimize the SSCHA free energy. Here we are representing phonons obtained with the SSCHA auxiliary force constants, those that we will use later to estimate superconducting critical temperature, instead of those derived from the Hessian of the SSCHA free energy (see Supp. Mat.~\cite{supp_mat}). There are three distinct sections in the phonon dispersion in both cases. The first section is formed by low-frequency phonon modes with dominant rubidium and phosphorus   character. These modes are not significantly affected by anharmonic and quantum effects. It is worth pointing out that phosphorus dominated modes are higher in frequency in the $Pm\bar{3}m$ phase compared to the $R3m$ phase, which is going to be important for the superconducting properties of these compounds. The second section is the middle-frequency hydrogen modes that are most affected by the SSCHA renormalization. The stiffening of the frequencies of these modes can exceed 1500 cm$^{-1}$ as in the case of the $Pm\bar{3}m$ phase. In the $Pm\bar{3}m$ phase, these phonon modes, as we have already discussed, are responsible for the phase transition. The third section is composed of high-frequency hydrogen modes that are not very affected by anharmonicity in the $Pm\bar{3}m$ phase and are somewhat stiffened in the $R3m$ phase.

\begin{figure}
    \centering
    \includegraphics[width=0.5\textwidth]{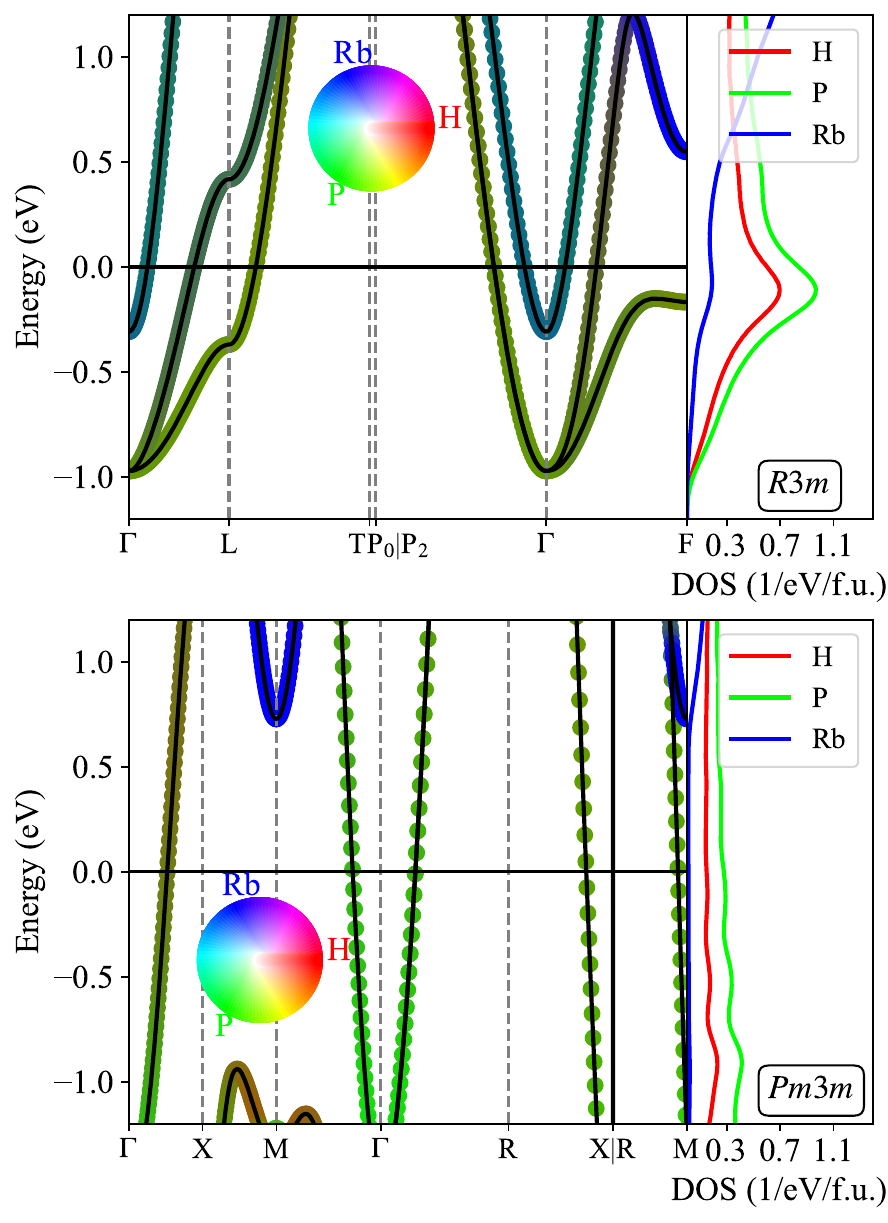}
    \caption{The electronic band structure of (a) the $R3m$ phase at 0 GPa and (b) the $Pm\bar{3}m$ phase at 30 GPa. The color of the points represents the projection of the state on the constituent atoms. The side plot shows the atom-projected electronic DOS.}
    \label{fig:electronic_bands}
\end{figure}

The electronic band structure of the $R3m$ phase for the SSCHA structure at 0 GPa has three bands crossing the Fermi level as shown in Fig.~\ref{fig:electronic_bands}, forming three concentric Fermi surfaces. 
The color of the points represents the character of the state determined by projecting the electronic wave function onto atomic orbitals. The states at the Fermi level coming from the two outer bands mainly have phosphorus and hydrogen mixed character, without separating into different bands. The third band that is closest to the Brillouin zone center is mostly of Rb character and has the lowest density of states (DOS). The Van Hove singularity gives rise to a distinct peak in the electronic DOS, located just 200 meV below the Fermi level. On the other hand, the electronic band structure of the $Pm\bar{3}m$ phase at 30 GPa is fairly simple, with a single band crossing the Fermi level with a large Fermi velocity  (see Fig.~\ref{fig:electronic_bands} (b)). In this case, there are virtually no Rb states at the Fermi level, while H and P states are still very mixed.


The large hydrogen content, large electron-phonon coupling, and the overall similarity with H$_3$S strongly suggest that RbPH$_3$ should be a good superconductor with a high \Tc. 
In the side plot of Fig.~\ref{fig:phonons} (a) we can see the Eliashberg spectral function of $R3m$ phase averaged over the Fermi surface and compared to the phonon density of states. We can see that most of the electron-phonon coupling comes from the hydrogen phonon modes, mainly in the middle-frequency range, while phosphorus and rubidium-dominated phonon modes contribute much less. It is interesting to remark that the most anharmonic modes are the ones that couple more strongly to the electrons, highlighting how anharmonicity stabilizes vibrational modes softened by a particularly strong electron-phonon interaction.
Using the commonly reported value of the reduced Coulomb potential $\mu^* = 0.14$ in isotropic Migdal-Eliashberg equations, we obtain a superconducting critical temperature estimate of 126 K for the $R3m$ phase at 0 GPa. However, if $\mu^*$ is calculated using Model-Anderson's formula and the screened Coulomb interaction is estimated in the random phase approximation (RPA)~\cite{Pellegrini_Minimal2022}, $\mu^*$ becomes 0.26, which reduces the calculated critical temperature (see Supp. Mat.~\cite{supp_mat}). The large value of the reduced Coulomb potential is a consequence of the large Matsubara cutoff used in our isotropic Migdal-Eliashberg equations and not a peculiarity of the electronic interaction of the RbPH$_3$ system. Indeed, the screened Coulomb interaction $\mu$ in the RPA takes a very common value of 0.39. 
Finally, we can use a purely {\it ab initio} approach in which the Coulomb interaction is computed and included without the arbitrary renormalization process as in the Morel-Anderson theory (see Supp. Mat.~\cite{supp_mat})
via the Eliashberg approach from Refs.~\cite{Pellegrini_Minimal2022,Sanna_Eliashberg_JPSJ2018}. 
In this case, the energy and Matsubara integrations are both performed explicitly leading to a critical temperature of 90~K, which is increased to 94.3 K if multiband effects are considered.
The agreement between isotropic and multiband approaches is not surprising considering that most of the electron-phonon coupling is coming from a single electronic band (see Supp. Mat. \cite{supp_mat}). While all of the mentioned methods for computing critical temperature are quite different, they all give fairly similar T$_\mathrm{C}$ estimates, confirming RbPH$_3$ as a promising candidate for high-temperature superconductivity at ambient pressure.

Finally, we have calculated the superconducting properties of the $Pm\bar{3}m$ phase at 30 GPa. In the side plot of Fig.~\ref{fig:phonons} (b) we show the Eliashberg spectral function of this phase. The electron-phonon coupling constant is lower than in the $R3m$ phase. The reason for this is its almost three times lower electronic DOS at the Fermi level compared to the $R3m$ phase. Furthermore, in the $Pm\bar{3}m$ phase phosphorus phonon modes are higher in frequency, which leads to the reduction of the electron-phonon coupling given that a large part of the electron DOS is of phosphorus character. This results in a lower estimate of T$_{\mathrm{C}}$ of 87 K in the isotropic approximation of the Migdal-Eliashberg theory using $\mu^* = 0.14$. Using phonons from the Hessian of the SSCHA free energy instead of SSCHA auxiliary ones leads to an increased estimate of 103 K. This trend is expected due to the phonon frequency softening in the Hessian case. The true critical temperature should lie somewhere in between these two values (87--103 K) for this isotropic case with the use of $\mu^* = 0.14$~\cite{H}.

In conclusion, we predicted that RbPH$_3$ is a dynamically and kinetically stable system at ambient pressures with a superconducting critical temperature of around 100 K thanks to ionic quantum anharmonic effects. RbPH$_3$ can be synthesized at relatively low pressures, at 30 GPa, in the cubic perovskite $Pm\bar{3}m$ phase. This phase is expected to undergo a second-order phase transition to a distorted rhombohedral phase below 20 GPa.   Our results suggest that trivalent atoms doped with alkali materials can form hydrides with perovskite-like structures and strong covalent bonds, clearly resembling the electronic structure and bonding of the high-T$_{\mathrm{C}}$ H$_3$S. More importantly, our calculations underline that ionic quantum anharmonic effects can stabilize at ambient pressures high-T$_{\mathrm{C}}$ compounds that are missed in current high-throughput calculations.   



\bigskip

This work is supported by the European Research Council (ERC) under the European Unions Horizon 2020 research and innovation program (Grant Agreement No. 802533), the Spanish Ministry of Science and Innovation (Grant No. PID2022142861NA-I00), the Department of Education, Universities and Research of the Eusko Jaurlaritza and the University of the Basque Country UPV/EHU (Grant No. IT1527-22), and Simons Foundation through the Collaboration on New Frontiers in Superconductivity (Grant No. SFI-MPS-NFS-00006741-10). We acknowledge EuroHPC for granting us access to Lumi located in CSC’s data center in Kajaani, Finland, (Project ID EHPC-REG-2024R01-084) and to RES for giving us access to MareNostrum5, Spain, (Project ID FI-2024-2-0035). Technical and human support provided by DIPC Supercomputing Center
is gratefully acknowledged. The authors acknowledge enlightening discussions with the partners of the SuperC collaboration.

\bibliography{paper}

	\onecolumngrid
    \renewcommand{\figurename}{Supplementary Figure}
    \renewcommand{\tablename}{Supplementary Table}
    \setcounter{figure}{0}
	\newpage
	\newpage
	\newpage

\section*{Supplementary material for: Ambient pressure high temperature superconductivity in RbPH$_3$ facilitated by ionic anharmonicity}

\vspace{5cm}

\subsection{Computational details}

To construct the convex hull we performed global structure prediction using the Minima Hopping Method (MHM)~\cite{Goedecker2004,Amsler2010}. We carried out MHM runs for all the stoichiometries within the Rb$_x$P$_y$H$_z$ formula, with $1 \leq x,y \leq 6$,  $1 \leq z \leq 10$, and $x+y+z \leq 12$. This corresponds to a total of 158 unique chemical compositions. For a given stoichiometry, the initial geometries were obtained randomly, ensuring only that the minimal distance between the atoms was at least equal to the sum of the covalent radii. Each minima hopping run was repeated twice. The 12 lowest energy structures for each composition were then re-optimized. Independent MHM runs were conducted at both 0 and 25 GPa. Additionally, we included all relevant materials from the Alexandria~\cite{Alexandria1, Alexandria2}, Super-Hydra~\cite{SuperHydra}, and Deepmind~\cite{deepmind} databases. In total, this amounted to 258 unique compositions. All the geometry optimizations and total energy calculations were performed with the code VASP~\cite{Kresse1996,Kresse1996_1}, using the PBE exchange and correlation functional and the PAW parameters~\cite{paw,paw2} of VASP version 5.2. As a final step, the lowest energy structure for each composition was re-optimized using Quantum Espresso using ultrasoft pseudopotentials with an energy cutoff of 70 Ry for electronic wave functions and a $\mathbf{k}$-point density of 0.006$\pi$ \AA$^{-1}$.

Density functional theory (DFT) and density functional perturbation theory (DFPT) calculations with
the Perdew-Burke-Ernzerhof parametrization~\cite{PBE} for the generalized gradient approximation were performed using the Quantum Espresso software package~\cite{QE1, QE2, QE3}. Ions were represented using ultrasoft pseudopotentials generated by the ``atomic'' code. Electronic-wave functions were defined in the plane-wave basis with an energy cutoff of 70 Ry, while the energy cutoff for the charge density was 280 Ry. The \textbf{k}-point grids used to sample the electronic states were $24\times 24\times 24$ for the $Pm\bar{3}m$ phase and $26\times 26\times 26$ for the $R3m$ phase. Due to the metallic nature of these compounds, we used a simple Gaussian smearing for electronic states of 0.01 Ry for the self-consistent calculations.

Stochastic self-consistent harmonic approximation (SSCHA) calculations were done on $2\times 2\times 2$ and $3\times 3\times 3$ supercells for both $R3m$ and $Pm\bar{3}m$ phases. A stochastic sampling of stresses and forces for the structural relaxations was done on 200 configurations for $2\times 2\times 2$ supercells and 400 configurations for $3\times 3\times 3$ supercells in both phases. Convergence of the Hessian of the free energy calculations required 2000 configurations for $2\times 2\times 2$ supercells in both phases, while for $3\times 3\times 3$ supercells in $R3m/Pm\bar{3}m$ phase it required 6000/4000 configurations.

The electron-phonon matrix elements needed for the Migdal-Eliashberg calculations were done on $8\times 8\times 8$ $\mathbf{q}$ point grids for both phases. The $\mathbf{k}$ point sampling for the Fermi surface average of these electron-phonon matrix elements for the $R3m$ phase was $42\times 42\times 42$, while for $Pm\bar{3}m$ phase was $48\times 48\times 48$. Double delta averaging of electron-phonon matrix elements over the Fermi surface was done with 0.006 Ry Gaussian smearing for the $R3m$ phase and 0.008 Ry Gaussian smearing for the $Pm\bar{3}m$ phase. The critical temperature was calculated solving Migdal-Eliashberg equations using SSCHA auxiliary phonons (see below for more details on the Migdal-Eliashberg equations). 


\newpage
\subsection{Self-consistent harmonic approximation (SSCHA)}

To investigate the structural and vibrational properties of RbPH$_3$, we employed the stochastic self-consistent harmonic approximation (SSCHA). The SSCHA method~\cite{SSCHA1,SSCHA2,SSCHA3,SSCHA4} minimizes the total free energy of the system, which accounts for quantum zero-point motion and anharmonicity, by optimizing two variational parameters: the centroid positions and auxiliary force constants, which define the ionic wavefunction. This wavefunction is assumed to be a Gaussian. The centroids represent the average atomic positions (the means of the Gaussian wavefunctions), while the auxiliary force constants are related to their standard deviations. The dynamical matrices constructed from these auxiliary force constants provide improved estimates of the phonon frequencies, as they are renormalized by anharmonicity, making them in principle more accurate than those obtained from harmonic force constants alone. 
Both the centroids and the SSCHA auxiliary second-order force constants are obtained through the free energy minimization process. 
	
Building on this non-interacting variational SSCHA picture, we can extend the analysis by considering an interacting picture within a many-body formalism in a manner consistent with the SSCHA formalism~\cite{SSCHA4, TDSSCHA, TDSSCHALihm, Siciliano}. In this framework, the phonon Green’s function, $G _{\mu\mu'}(\mathbf{q}, \omega)$, can be written as:
	\begin{align*}
	G _{\mu\mu'}(\mathbf{q}, \Omega) = \left[\Omega ^2\delta_{\mu\mu'} - \stackrel{(2)}{D} _{\mu\mu'}(\mathbf{q}) - \Pi _{\mu\mu'}(\mathbf{q} ,\Omega)\right]^{-1}.
	\end{align*}
	Here $\stackrel{(2)}{D} _{\mu\mu'}(\mathbf{q})$ is the dynamical matrix constructed from the SSCHA auxiliary force constants and $\Pi _{\mu\mu'}(\mathbf{q},\omega)$ is phonon self-energy that depends on the SSCHA anharmonic force constants ($ \stackrel{(3)}{\boldsymbol{\mathcal{D}}}(\mathbf{q}), \stackrel{(4)}{\boldsymbol{\mathcal{D}}}(\mathbf{q})$):

    \begin{align}
	\boldsymbol{\Pi}(\mathbf{q}, \Omega) = \stackrel{(3)}{\boldsymbol{\mathcal{D}}}(\mathbf{q}):\boldsymbol{\Lambda}(\mathbf{q},\Omega):\left[ \boldsymbol{1} -  \stackrel{(4)}{\boldsymbol{\mathcal{D}}}(\mathbf{q}):\boldsymbol{\Lambda}(\mathbf{q}, \Omega)\right]^{-1}:\stackrel{(3)}{\boldsymbol{\mathcal{D}}}(\mathbf{q}). 
        \label{eq:self-energy}
	\end{align}
The double-dot product $\mathbf{X}:\mathbf{Y}$ indicates the contraction of the last two indices of $\mathbf{X}$ with the first two indices of $\mathbf{Y}$. If we denote the eigenvalues of the SSCHA auxiliary dynamical matrices as $\omega _{\mu}(\mathbf{q})$ and associated Bose-Einstein factors as $n_{\mu}(\mathbf{q})$, the above $\boldsymbol{\Lambda}(\mathbf{q}, \Omega)$ is given as:
		\begin{equation}
		\Lambda ^{\mu\mu'}(\mathbf{q}, \Omega) = \frac{1}{4\omega _{\mu}(\mathbf{q})\omega _{\mu'}(\mathbf{q})}\left[\frac{\left(\omega _{\mu}(\mathbf{q}) - \omega _{\mu'}(\mathbf{q})\right)\left(n_{\mu}(\mathbf{q}) - n_{\mu'}(\mathbf{q})\right)}{(\omega _{\mu}(\mathbf{q}) - \omega _{\mu'}(\mathbf{q}))^2 - \Omega ^2 + i\epsilon} - \frac{(\omega _{\mu}(\mathbf{q}) + \omega _{\mu'}(\mathbf{q}))(1 + n_{\mu}(\mathbf{q}) + n_{\mu'}(\mathbf{q}))}{(\omega _{\mu}(\mathbf{q}) + \omega _{\mu'}(\mathbf{q}))^2-\Omega ^2 + i\epsilon}\right]. 
		\label{eq:lambda}
		\end{equation}
The quantity $\Pi _{\mu\mu'}(\mathbf{q},\Omega)$ is not purely real and accounts for the realistic broadening of the phonon spectral functions. However, in the static limit ($\Omega \rightarrow 0$), the contributions from these terms become purely real and can be used to further renormalize the SSCHA second-order auxiliary force constants. The resulting force constants represent the Hessians of the total free energy, $G _{\mu\mu'}(\mathbf{q}, 0)$. If any of the eigenvalues of this Hessian are negative, the structure is unstable. These renormalized force constants can also be used to describe the vibrational properties of the material. 

\begin{figure}[h!]
    \centering
    \includegraphics[width=0.5\linewidth]{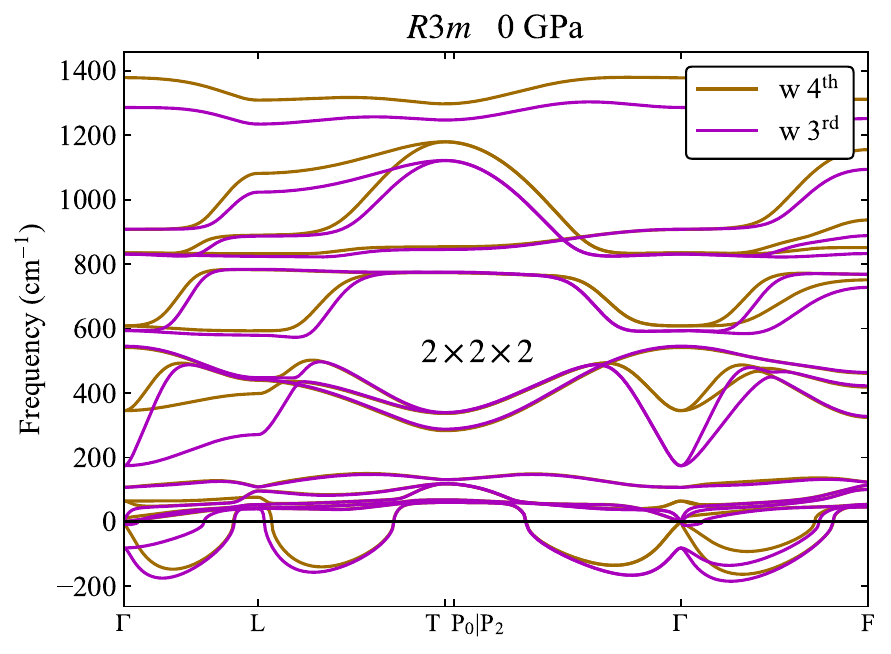}
    \caption{Phonon band structure of $R3m$ phase obtained using Hessian force constants with inclusion of 3$^{\mathrm{rd}}$ and 4$^\mathrm{th}$ order anharmonicity via Supp. Eq.~\ref{eq:self-energy}.}
    \label{fig:compare_hessians}
\end{figure}

In our calculations, we can calculate the Hessian frequencies with inclusion of the third and fourth-order force constants. However, in systems with many atoms (like $3\times 3\times 3$ supercell in our case) the calculation of the fourth-order correction becomes impossible due to a large RAM memory requirement and, in this cases, we set $\stackrel{(4)}{\boldsymbol{\mathcal{D}}}(\mathbf{q})=0$ in Supp. Eq. \eqref{eq:self-energy}. This is unfortunate as the third-order correction is negative definite, meaning it will always lead to lower phonon frequencies and can lead to the underestimation of the stability of the studied system. This is what we see in RbPH$_3$, where including only third-order anharmonicity leads to unstable Hessian frequencies, which become positive with the inclusion of $\stackrel{(4)}{\boldsymbol{\mathcal{D}}}(\mathbf{q})$ in the $2\times 2\times 2$ supercell case (see Supp. Fig.~\ref{fig:compare_hessians}) at the points commensurate with the supercell.

\newpage
\subsection{Analysis of SSCHA results}

Here we would like to describe in detail the different SSCHA calculations we performed. We performed SSCHA calculations in $2\times 2\times 2$ and $3\times 3\times 3$ supercells for both phases. In the case of $2\times 2\times 2$ supercell, we are able to include $\stackrel{(4)}{\boldsymbol{\mathcal{D}}}(\mathbf{q})$ in the calculations of the Hessian. On the other hand, including 4$^\mathrm{th}$ order anharmonicity for the case of $3\times 3\times 3$ supercell is too computationally expensive and the Hessian calculation only accounts for third-order anharmonicity, e.g. $\stackrel{(3)}{\boldsymbol{\mathcal{D}}}(\mathbf{q})$.

\begin{figure}
    \centering
    \includegraphics[width=0.45\linewidth]{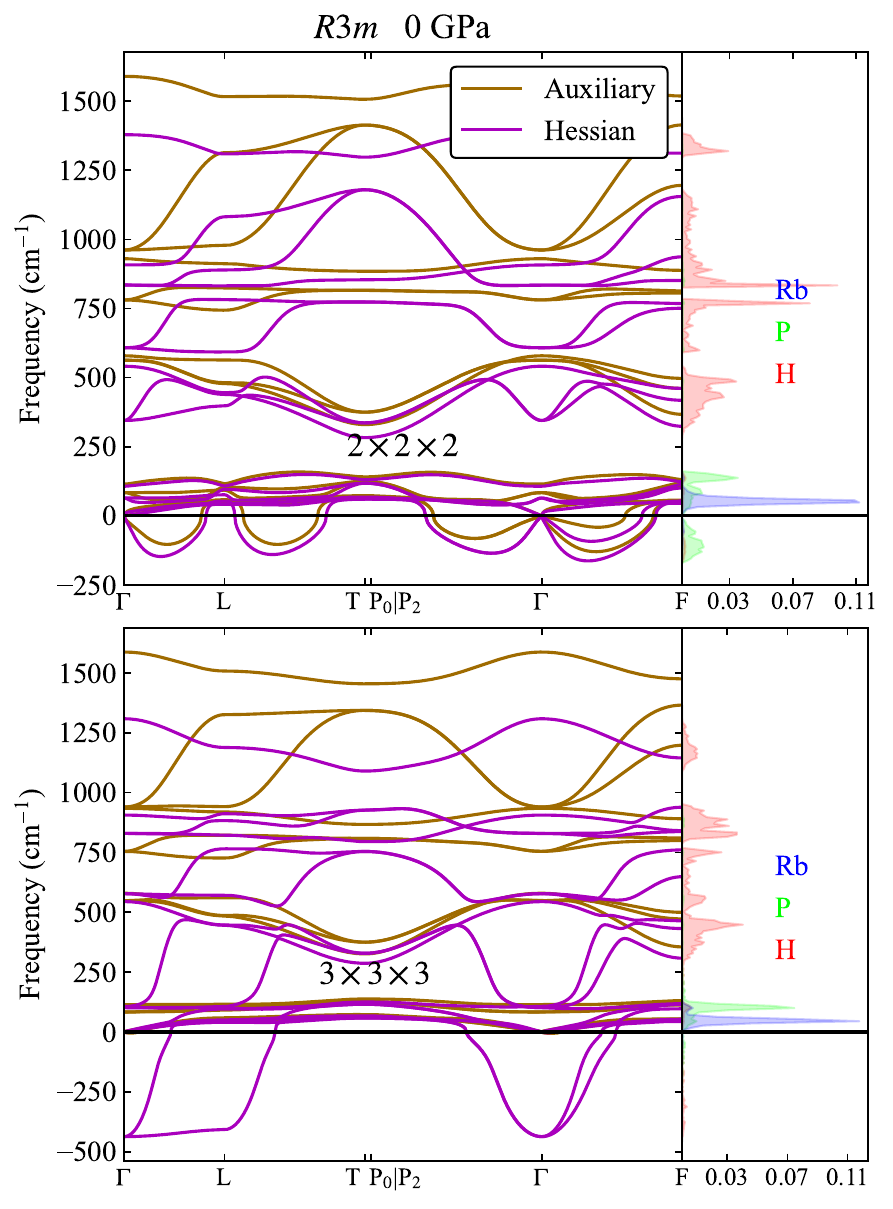}
    \includegraphics[width=0.45\linewidth]{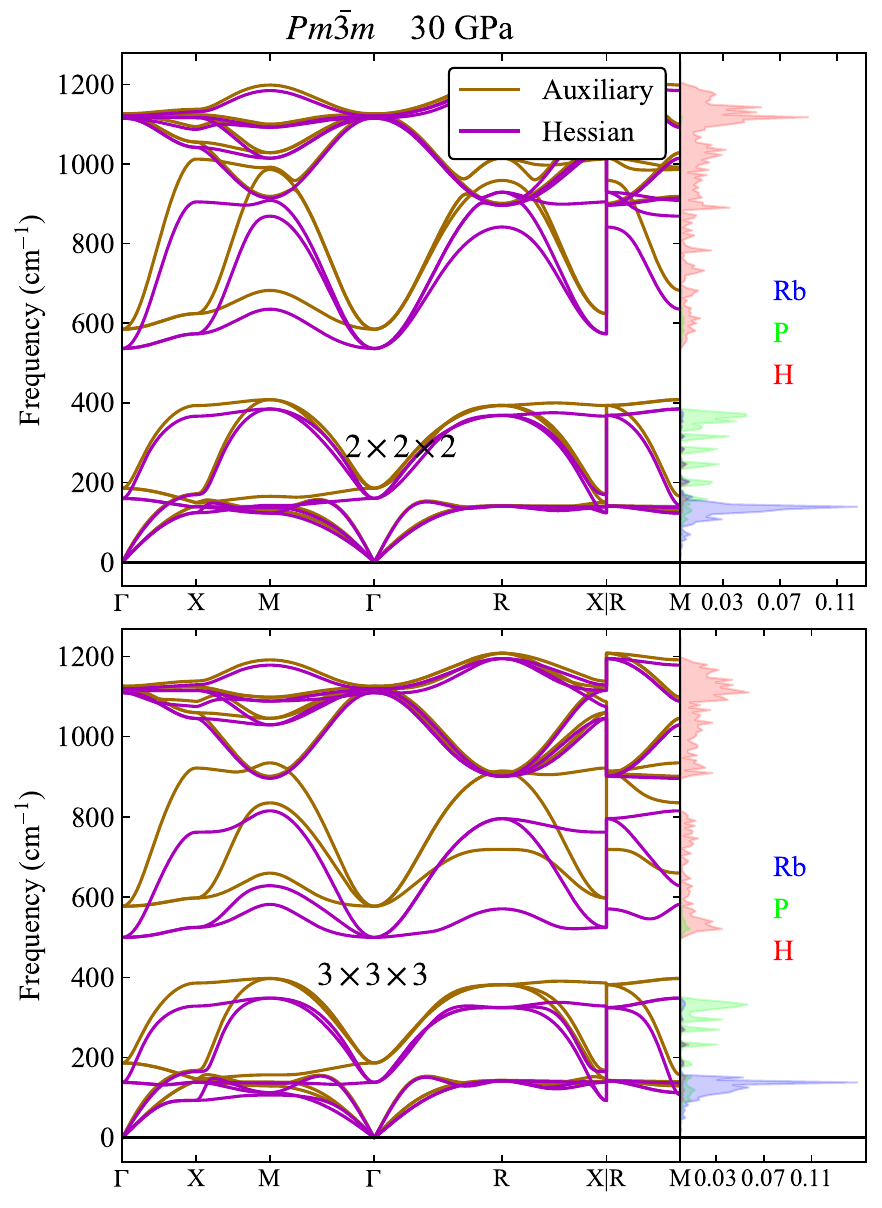}
    \caption{Phonon band structure of $R3m$ and $Pm\bar{3}m$ phase of RbPH$_3$ calculated with Hessian (including only 3$^{\mathrm{rd}}$ order anharmonicity) and auxiliary SSCHA force constants calculated on $2\times 2\times 2$ and $3\times 3\times 3$ supercells. Side plots show the phonon density of states for Hessian phonons. The phonon density of states for auxiliary phonons is shown in the main part.}
    \label{fig:sscha_phonons}
\end{figure}

In the $Pm\bar{3}m$ phase at 30 GPa the 4$^\mathrm{th}$ order anharmonicity does not play a significant role. It somewhat stiffens the phonon modes across the Brillouin zone as we would expect. Also in this phase, the interpolation of phonons does not pose a great challenge and even a $2\times 2\times 2$ $\mathbf{q}$ grid is enough for a stable interpolation.

In the $R3m$ phase at 0 GPa the 4$^\mathrm{th}$ order anharmonicity plays a crucial role. As we can see, in the commensurate $\mathbf{q}$ points for the $2\times 2\times 2$ grid (most notably the $\Gamma$ point) we have stable phonons when we include the $\stackrel{(4)}{\boldsymbol{\mathcal{D}}}(\mathbf{q})$ (see Supp. Fig.~\ref{fig:compare_hessians}). If we do not include it, as is the case in $3\times 3\times 3$ supercell, we see unstable phonon modes in the commensurate $\mathbf{q}$ points. This also happens in the $2\times 2\times 2$ supercell if we only include the third-order anharmonicity in the calculation of Hessian. Another issue with the $R3m$ phase is that we have unstable phonon interpolation in the $2\times 2\times 2$ supercell. Even in the case of auxiliary phonons, we see unstable phonon modes at interpolated $\mathbf{q}$ points, something that should not be possible. This is signaling that the $2\times 2\times 2$ supercell is not large enough to converge results. We do not have the same issue with $3\times 3\times 3$ supercell. However, in this case, we do not have the stable hessian modes since we can not add 4$^\mathrm{th}$ order anharmonicity.

\newpage
\subsection{RbPH$_3$ SSCHA structure}

Here we give the structural parameters of $R3m$ and $Pm\bar{3}m$ phase of RbPH$_3$ in a rhombohedral setting calculated at the BOES level and within the SSCHA. $a$ is the lattice constant, $\theta$ is the angle between lattice vectors. Wyckoff positions in this setting are given as: Rb (0,0,0); P (z,z,z); H (x,y,x); H (x,x,y); H (y,x,x).

\begin{table}[h!]
\begin{tabular}{|cc|c|c|c|c|c|}
\hline
\multicolumn{2}{|l|}{}                                               & $a$   & $\theta$ & P(z)  & H(x)  & H(y)  \\ \hline
\multicolumn{1}{|c|}{\multirow{2}{*}{0 GPa}}  & $R3m$ (BOES)         & 4.233 & 90.777   & 0.469 & 0.480 & 0.832 \\ \cline{2-7} 
\multicolumn{1}{|c|}{}                        & $R3m$ (SSCHA)        & 4.297 & 91.071   & 0.473 & 0.483 & 0.835 \\ \hline
\multicolumn{1}{|c|}{\multirow{2}{*}{30 GPa}} & $Pm\bar{3}m$ (BOES)  & 3.514 & 90.000   & 0.500 & 0.500 & 0.000 \\ \cline{2-7} 
\multicolumn{1}{|c|}{}                        & $Pm\bar{3}m$ (SSCHA) & 3.542 & 90.000   & 0.500 & 0.500 & 0.000 \\ \hline
\end{tabular}
\end{table}

\begin{figure}
    \centering
    \includegraphics[width=0.5\linewidth]{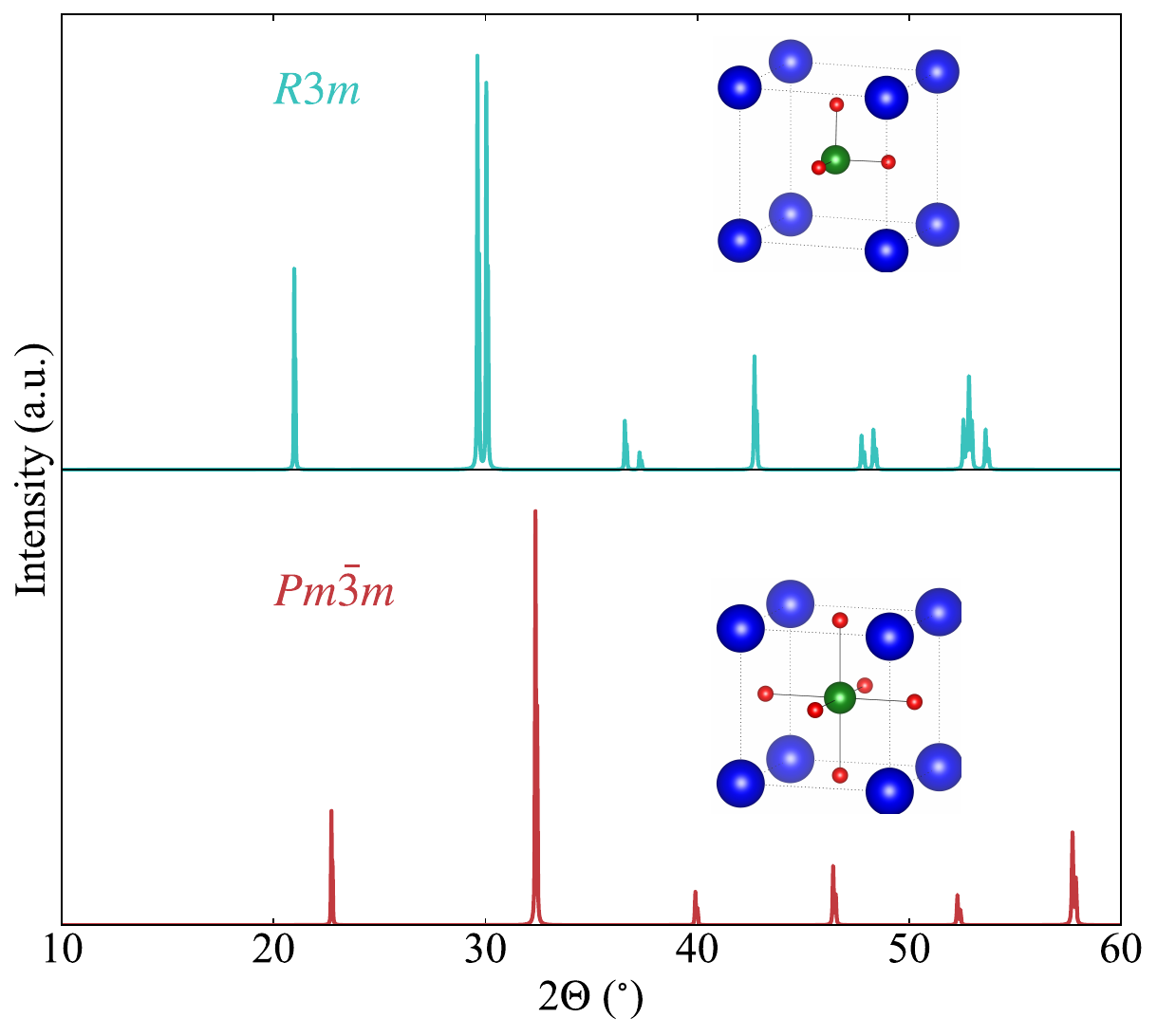}
    \caption{XRD pattern of $R3m$ and $Pm\bar{3}m$ phases of RbPH$_3$ at 0 GPa as determined from Born-Oppenheimer energy surface.}
    \label{fig:xrds}
\end{figure}
\newpage

\subsection{Multiband Migdal-Elishberg equations}\label{ME:multiband}

Since RbPH$_3$ in the $R3m$ phase has three nicely separated Fermi surfaces we decided to implement multiband Migdal-Elishberg equations in order to estimate the influence of an anisotropic Fermi surface on the superconducting properties. In the multiband approach we have to solve the following self-consistent set of equations:
\begin{align}
     Z_i (i\omega _n) &= 1+ \frac{\pi T}{\omega _n}\sum _{m,j}\frac{\omega _mZ_j(i\omega _m)\lambda _{ij}(i\omega _n, i\omega _m)}{\sqrt{(\omega _m Z_j(i\omega _m))^2  + \phi ^2 _j (0, i\omega _m)}} \\
     \phi ^{ph} _i (i\omega _n) &= \pi T \sum _{m,j} \frac{\lambda _{ij}(i\omega _n, i\omega _m)\phi _j (0, i\omega _m)}{\sqrt{(\omega _m Z_j(i\omega _m))^2  + \phi ^2 _j (0, i\omega _m)}} \\
     \phi _i ^{c} (\varepsilon) = -\sum _{j}\int \mathrm{d}\varepsilon ' W_{ij}(\varepsilon, \varepsilon ')N_j(\varepsilon)&\left\{\frac{1}{2}\frac{\tanh{\frac{1}{2T}\sqrt{{\varepsilon '} ^2 + (\phi _j ^{c}(\varepsilon '))^2}}}{\sqrt{{\varepsilon '} ^2 + (\phi _j ^{c}(\varepsilon '))^2}} + T\sum_{m} \left[\frac{\phi _j (\varepsilon ',i\omega _m)}{\Theta _j (\varepsilon ', i\omega _m)} - \frac{\phi _j (\varepsilon ',i\omega _m)}{\omega _m ^2 + \varepsilon ^2 + (\phi _j ^{c}(\varepsilon '))^2}\right]\right\} \label{eq:full}\\
    \phi _i (\varepsilon, i\omega _n) &= \phi ^{ph} _i (i\omega _n) + \phi ^c _i (\varepsilon) \\
     \Theta _i (\varepsilon, i\omega _n) &= (\omega _nZ_i (\omega _n))^2 + \varepsilon ^2 + \phi ^2 _i (\varepsilon, i\omega _n)
\end{align}
Here, $\omega _n$ are Matsubara frequencies, $Z_i (i\omega _n)$ is the mass renormalization function for band $i$, while $\phi _i (i\omega _n)$ is the superconducting order parameter for the same band. $W_{ij}(\varepsilon, \varepsilon')$ is the energy surface averaged screened Coulomb interaction (see next section for more details). In case we take this value at the Fermi level ($\varepsilon = 0$) and disregard energy variation of the electronic density of states we can express the Coulomb part of the order parameter through renormalized Coulomb interaction $\mu _{ij}^*$:
\begin{align}
    \phi _i ^{c}  = -\pi T\sum _{j,\omega _m} \frac{\mu _{ij}^*\phi _j (i\omega _m)}{(\omega _m Z_j (i\omega _m))^2 + (\phi _j (i\omega _m))^2}.\label{eq:renorm}
\end{align}

We have performed calculations with both considering fully Supp. Eq.~\ref{eq:full}, i.e. without renormalization, and with the simplified approach via Supp. Eq.~\ref{eq:renorm}. Finally, if we omit band indices and sums over bands we recover the standard isotropic solution to the Migdal-Eliashberg equations. 

In the multiband case, the electron-phonon coupling constant $\lambda$ is a $N\times N$ matrix where $N$ is the number of bands that are crossing the Fermi surface:
\begin{align*}
    \lambda _{ij}(i\omega _n, i\omega _m) = 2\int _0 ^{\infty}\frac{\Omega}{\Omega ^2 + (\omega _n - \omega _m)^2}\alpha ^2F _{ij}(\Omega)\mathrm{d}\Omega .
\end{align*}
Analogously $\alpha ^2F _{ij}(\Omega)$ is the $N\times N$ Elishberg spectral function matrix:
\begin{align*}
    \alpha ^2F _{ij}(\Omega) = \frac{1}{N_\mathbf{q}}\sum_{a,b,\mathbf{q}} \Delta ^{ab} _{ij} (\mathbf{q})B^{ab}(\mathbf{q},\Omega),
\end{align*}
where $a,b$ are compact Cartesian and atom indices, $\mathbf{q}$ is the phonon wave vector, $B^{ab}(\mathbf{q},\Omega)$ is the phonon spectral function and $\Delta ^{ab} _{ij} (\mathbf{q})$ is the band resolved electron-phonon matrix element averaged over Fermi surface:
\begin{align*}
    \Delta ^{ab} _{ij} (\mathbf{q}) = \frac{1}{N^F _{i}N_\mathbf{k}}\sum _\mathbf{k} d^a _{i\mathbf{k},j\mathbf{k}+\mathbf{q}}d^b _{j\mathbf{k}+\mathbf{q}, i\mathbf{k}}\delta (\epsilon _{i\mathbf{k}} - \epsilon _F)\delta (\epsilon _{j\mathbf{k} + \mathbf{q}} - \epsilon _F).
\end{align*}
Here $\mathbf{k}$ is the wave vector of electronic state, $N^F _i$ is the electronic density of states at the Fermi level of band $i$, $\epsilon _{i\mathbf{k}}$ is the energy of electronic state of wave vector $\mathbf{k}$ and band index $i$, and $\epsilon _F$ is the Fermi level. $d^a _{i\mathbf{k},j\mathbf{k}+\mathbf{q}}$ is the deformation potential: $d^a _{i\mathbf{k},j\mathbf{k}+\mathbf{q}} = \langle i\mathbf{k}|\frac{\delta V}{\delta u^a(\mathbf{q})}|j\mathbf{k}+\mathbf{q}\rangle$. It is easy to show that the isotropic Elishberg spectral function is obtained via ($N_F$ is the total electronic density of states at the Fermi level):
\begin{align*}
    \alpha ^2F(\Omega) = \frac{\sum_{i,j}N^F _i \alpha ^2F _{ij}(\omega)}{N_F}.
\end{align*}

\begin{figure}
    \centering
    \includegraphics[width=0.9\linewidth]{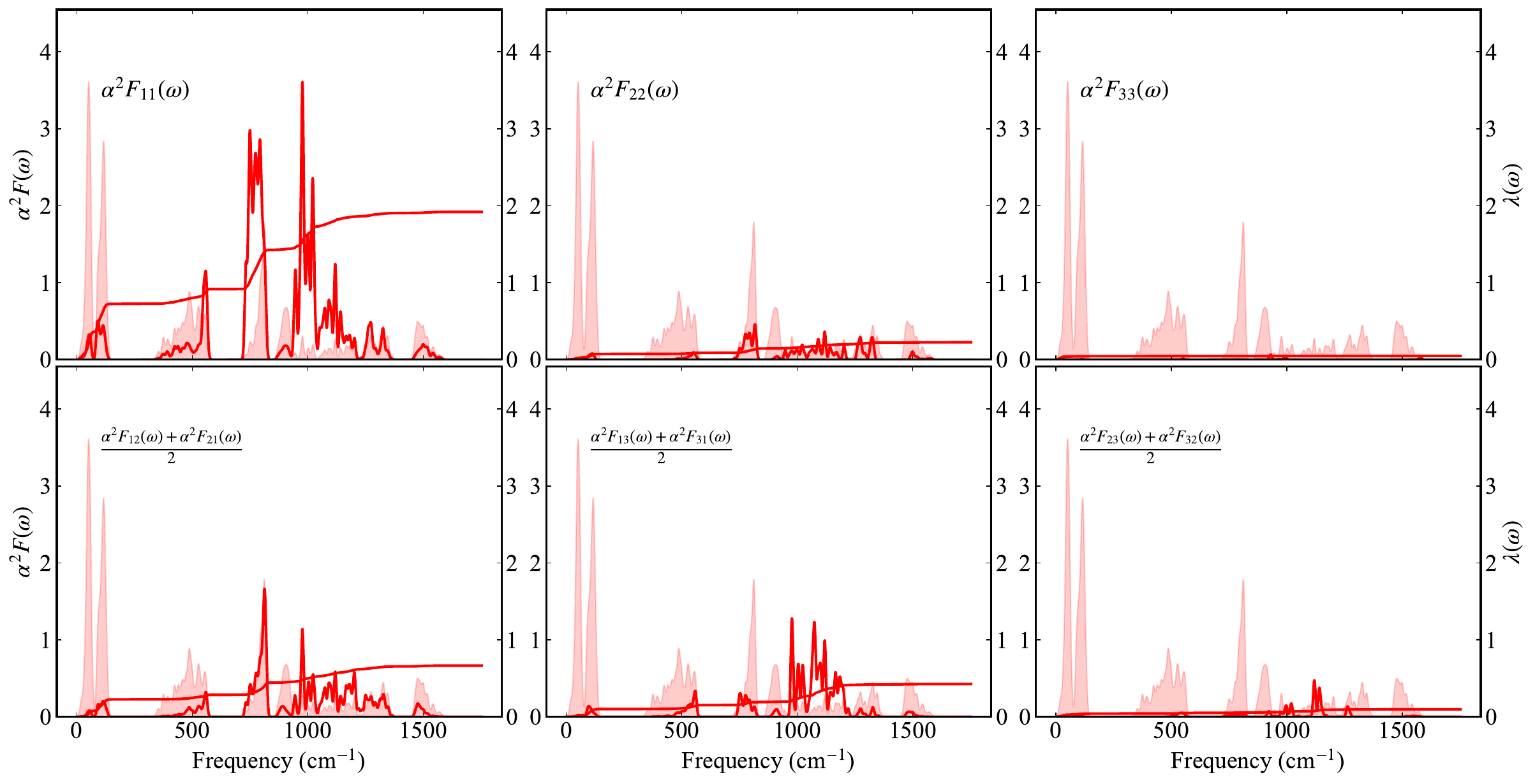}
    \caption{Band resolved Eliashberg spectral function $\alpha ^2F(\omega)$ and electron-phonon coupling constant ($\lambda$). The background is the phonon density of states scaled.}
    \label{fig:a2fs}
\end{figure}

In Supp. Fig.~\ref{fig:a2fs} we are showing band-resolved Eliashberg spectral functions and electron-phonon coupling constants. Most of the electron-phonon coupling comes from the band with the highest density of states. This is probably why multiband and isotropic approximations give similar results (see below).

\newpage
\subsection{Coulomb interaction}


The Coulomb interaction in conventional superconductors reduces the binding within the Cooper pairs. We denote by $W_{k,k'}$ the matrix elements of the screened Coulomb interaction with respect to the Kohn-Sham orbitals, 
\begin{equation}
W_{k,k'}=4\pi\sum_{\boldsymbol{G},\boldsymbol{G}'}\epsilon^{-1}_{\boldsymbol{G}\boldsymbol{G}'}(\boldsymbol{q})
\times \frac{\langle k'|e^{-i(\boldsymbol{q}+\boldsymbol{G})\cdot \boldsymbol{r}}|k\rangle \langle k|e^{i(\boldsymbol{q}+\boldsymbol{G}')\cdot \boldsymbol{r}}|k'\rangle}{|\boldsymbol{q}+\boldsymbol{G}| |\boldsymbol{q}+\boldsymbol{G}'|},\label{eq:Wkkp}
\end{equation}
where $k=(\boldsymbol{k},n)$ is a combined momentum and band index, $\bf{G}$ are reciprocal lattice vectors and we have assumed the static approximation for the dielectric matrix $\epsilon^{-1}_{\boldsymbol{G}\boldsymbol{G}'}(\boldsymbol{q})$.

In the isotropic approximation~\cite{Massidda_SUST_CoulombSCDFT_2009, mu1,Sanna_Eliashberg_JPSJ2018,   SPG_EliashbergSCDFT_PRL2020,PellegriniKukkonenSanna_beyondRPA_PRB2023} Supp. Eq.~\eqref{eq:Wkkp} is averaged over surfaces of constant energy ($\varepsilon$) in $k$-space, i.e.:
\begin{equation}
W(\varepsilon,\varepsilon')=\frac{1}{N(\varepsilon)N(\varepsilon')} \sum_{k,k'} W_{k,k'}\delta(\varepsilon_k-\varepsilon)\delta(\varepsilon_{k'}-\varepsilon'),\label{eq:Waverage}
\end{equation} 
where $N(\varepsilon)=\sum_k \delta(\varepsilon_k-\varepsilon)$ is the density of states.
By experience, the value of $T_c$ is mostly affected by the  Coulomb repulsion at the Fermi level. This is usually expressed by the parameter $\mu$, 
\begin{equation}
    \mu=N(0)W(0,0),\label{eq:mu}
\end{equation}
defined as the average on the Fermi surface of the electron-electron matrix elements times the density of states at the Fermi level.
In simple superconductors with a good metallic screening 
$W$  can be computed reasonably accurately in the random phase approximation~\cite{PellegriniKukkonenSanna_beyondRPA_PRB2023} and  $\mu$ typically acquires values within the range 0.2-0.4. 

In the present case, we compute $\mu=0.39$ which, owing to the covalent nature of this compound and its relatively high density of states at the Fermi level, is on the high end of the typical interval. 
The calculation was done with the elk-FLAPW code~\cite{ElkCode}. The primary convergence parameters for an accurate calculation of $W$ are the k and q-point grids (here converged at $8\times8\times8$) and the radius of ${\bf G}$ vectors integrated in Eq.~\eqref{eq:Wkkp} converged here to $|{\bf G}|=3.0$. 

\begin{figure}
    \centering
    \includegraphics[width=0.4\linewidth]{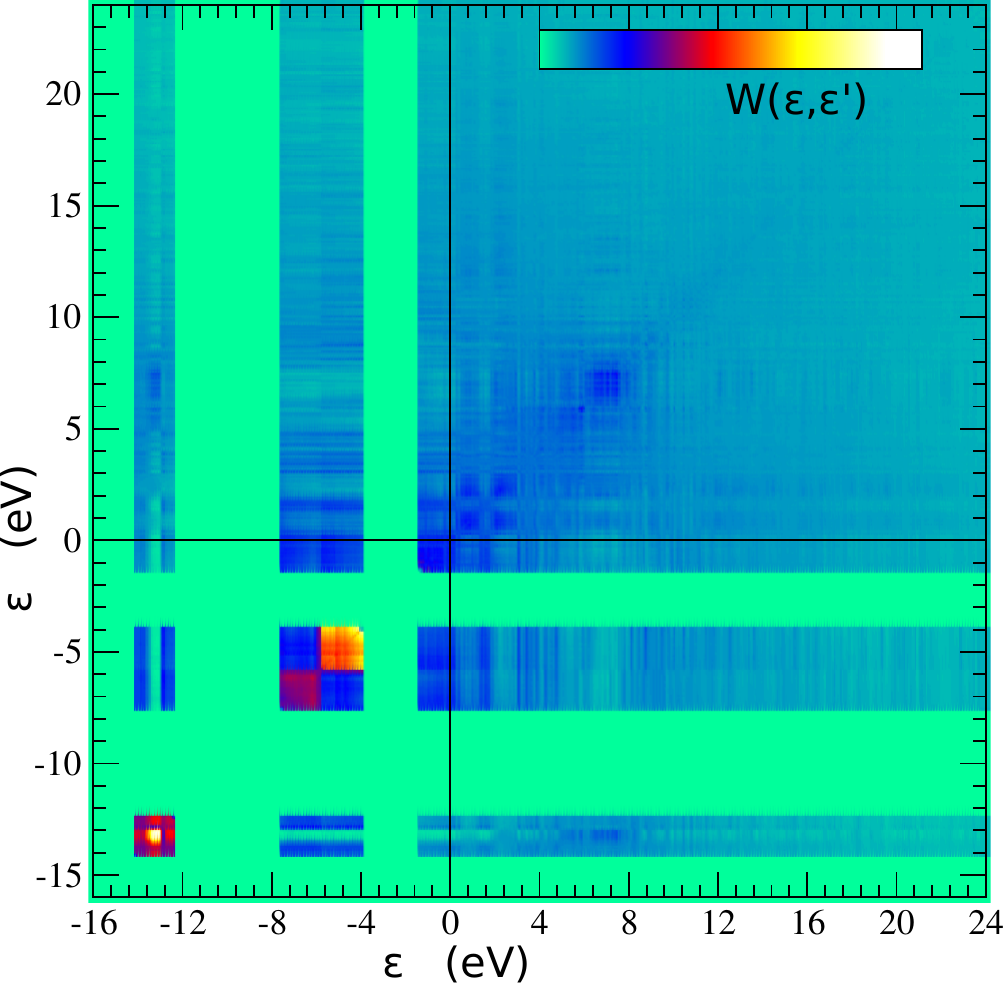}
    \caption{Coulomb interaction as a function of the band energies $W(\varepsilon,\varepsilon')$. The function is computed in the random phase approximation for the dielectric screening.}
    \label{fig:W}
\end{figure}

In the classical multiband Migdal-Eliashberg approach of Sec.~\ref{ME:multiband} we are using the renormalized value of Coulomb potential $\mu ^*$. We can estimate this parameter using the Morel-Anderson theory:
\begin{equation}\label{eq:MorelAnderson}
\mu^*=\frac{\mu}{1+\mu\ln\left(\frac{E_F}{\omega_c}\right)},
\end{equation}
where $\omega_c$ is the renormalization cutoff, which is arbitrary except for being larger than the Debye energy, $E_\mathrm{F}$ is a parameter of the order of the valence bandwidth. We estimated the value of $E_\mathrm{F}$ to be 7.5 eV and chose $\omega _c$ to be 10 times the maximum phonon frequency (1.97 eV) which results in $\mu ^* = 0.26$. This value is considerably larger than the ones usually used in the literature and the reason for this is the large value of $\omega _c$. Hydride materials have exceptionally large Debye frequencies which makes renormalized Coulomb interaction larger than in simple metals. 

In the table below we summarize the results for different ways of treating electron-phonon and Coulomb effects in the $R3m$ phase of RbPH$_3$.
The ``Full Coulomb'' column corresponds to calculations performed with Full Migdal Eliashberg approach~\cite{Sanna_Eliashberg_JPSJ2018,Pellegrini_Minimal2022} (see Eq.~\ref{eq:full}), where Coulomb interactions are included in the entire energy range (using the function $W(\varepsilon,\varepsilon')$ as defined in Eq.~\ref{eq:Waverage}). 
In this way the Coulomb renormalization is intrinsically included in the solution of a full band-width set of Eliashberg equations. Considering that phonon anisotropy is very weak, we are using an isotropic form $W(\varepsilon,\varepsilon')$. Although a generic band resolved  $W_{ij}(\varepsilon,\varepsilon')$ can be easily computed by adjusting the summation domains of Eq.~\ref{eq:Waverage}. 

\begin{table}[h!]
\begin{tabular}{|c|c|c|c|c|}
\hline
 method & $\mu^*$ = 0.14 (Eq.~\ref{eq:renorm}) & $\mu^*$ = 0.26 (Eq.~\ref{eq:renorm})& Full Coulomb (Eq.~\ref{eq:full}) & Full Coulomb multiband (Eq.~\ref{eq:renorm}) \\ \hline
T$_\mathrm{C}$      & 126 K          & 101 K          & 90 K          & 94 K     \\ \hline
\end{tabular}
\caption{Estimated superconducting critical temperature of $R3m$ phase of RbPH$_3$ at 0 GPa using different approaches.}
\end{table}





\end{document}